\documentclass[aps,prb,showpacs,preprintnumbers,amsmath,%
amssymb,preprint]{revtex4}
\begin{document}  
\sloppy         
\title{Non-Fickian Interdiffusion of Dynamically Asymmetric Species: 
A Molecular Dynamics Study}
\author{Jacqueline Yaneva$^{1,2}$, 
Burkhard D\"{u}nweg$^2$ and Andrey Milchev$^{1,2}$}
\affiliation{
$^1$Institute for Physical Chemistry \\
Bulgarian Academy of Sciences \\
1113 Sofia, Bulgaria \\
$^2$ Max-Planck-Institut f\"{u}r Polymerforschung \\
Ackermannweg 10 \\ D-55128 Mainz, Germany\\
}
\date{\today}
\begin{abstract}            
We use Molecular Dynamics combined with Dissipative Particle Dynamics
to construct a model of a binary mixture where the two species differ
only in their dynamic properties (friction coefficients). For an
asymmetric mixture of slow and fast particles we study the
interdiffusion process. The relaxation of the composition profile is
investigated in terms of its Fourier coefficients. While for weak
asymmetry we observe Fickian behavior, a strongly asymmetric system
exhibits clear indications of anomalous diffusion, which occurs in a
crossover region between the Cases I (Fickian) and II (sharp front
moving with constant velocity), and is close to the Case II
limit.
\end{abstract}
\pacs{66.10.-x,66.10.Cb,02.70.Ns}

\maketitle

\section{INTRODUCTION}
\label{INT}

The study of anomalous diffusion phenomena in polymeric materials has
been of interest for decades. One particular instance of non-Fickian
transport is Case II diffusion\cite{Crank}, which is usually
observed in glassy polymers subjected to penetration by a
low-molecular weight solvent. The most characteristic feature of this
phenomenon is the development of a sharp front in the concentration
profile, which advances {\em linearly} in time, and ahead of which the
penetrant concentration is very low. The dynamics is therefore
characterized by a single parameter, the velocity of the front. In
contrast, Fickian (Case I) diffusion is described by the diffusion
coefficient and a corresponding scaling of position (and amount of
sorbed penetrant) with $t^{1/2}$.

Case II diffusion has been intensively investigated
experimentally\cite{Hartley,Kwei,Kramer,Russell} and
theoretically\cite{Peterlin,Wang,KWZ,Argon,Lee,Elafif,Witelski,Edwards,%
TW80,RossiPRE,deGennes95,RossiMac,Taylor00},
and various microscopic and phenomenological models have been proposed
to explain the observed features. Among the most popular are the
models of Thomas and Windle\cite{TW80} (TW), and of Rossi {\it et
al.}\cite{RossiPRE,deGennes95,RossiMac}, which was recently
extended by Qian and Taylor\cite{Taylor00}.

Regardless of the details of the various models, the development of
the linearly advancing front is rather easy to
understand\cite{Kramer}. The two decisive preconditions are (i) the
existence of a strong disparity in the mobilities of the two pure
species (glassy matrix vs. penetrant), and (ii) the ``plasticizing
effect'', i.~e. a strong enhancement of the mobility of the slow
species (constituent of the glassy matrix) if its molecules
are surrounded by those of the fast species (penetrant). As soon as
the slow molecules are plasticized at the front, they quickly make
room for the penetrant, which in turn is rapidly transported into the
thus-opened free volume. This results in a spatially constant
concentration profile behind the front. The rate of plasticization
(matrix dissolution) is determined by the penetrant concentration at
the front, and therefore remains constant. This gives rise to a
constant-velocity front.

One should note that this generic scenario does not necessarily
require a polymeric matrix -- the ``dynamic contrast'' between the two
species is not specific to polymers. In fact, quite similar behavior
has been observed in the dissolution of low molecular weight silicate
glasses\cite{Perera}, where the constant rate of dissolution was
attributed to a constant rate of chemical reaction (hydration) at the
surface. Nevertheless, most studies are concerned with polymer
systems, where the phenomenon is of particular practical relevance.
There are {\em additional} polymer-specific effects beyond simple Case
II behavior, like entanglements and the resulting stresses in the
matrix\cite{RossiPRE}, which are technologically important.

For such reasons, one would like to be able to simulate the phenomenon
on the computer in order to gain insight into the underlying molecular
mechanisms. A recent attempt has been undertaken by M. Tsige and
G. S. Grest\cite{Grest04,Grest04_1} who investigated a coarse-grained
polymer-solvent system by means of Molecular Dynamics and grand
canonical Monte Carlo simulation. However, these authors did not
observe any deviation from Fickian behavior, probably because the
simulations failed to reach the necessary separation of time scales,
even though the polymers were able to vitrify. 

In the present study we attempt a quite different simulational
approach in order to create the necessary dynamic disparity.  As
outlined above, one expects that Case II behavior (i.~e., a linearly
propagating sharp front, and a flat concentration profile behind the
front) should occur as soon as this disparity is sufficiently strong,
and one takes into account the ``plasticizing'' effect -- even if the
model does not exhibit any typical polymer features.  As our results
indicate (see below), our model, though very simple, seems to grasp
the main features on non-Fickian diffusion at least qualitatively. We
believe that a combination of our approach with a polymer model would
prove promising in revealing the molecular aspects of anomalous
diffusion in glassy polymers. The purpose of our study is hence
twofold: On the one hand, we wish to further corroborate the view that
dynamic disparity is indeed the decisive ingredient for Case II
behavior. On the other hand, we propose a methodological advancement
in the microscopic simulation of such phenomena.

The simplest model in this spirit is a binary system of particles
which are completely identical with respect to their static properties
(i.~e. their interaction potential), but differ in their dynamic
properties. One possibility to do this is to run a Molecular Dynamics
simulation of a binary mixture whose species are identical except for
their masses. This approach has, for instance, been pursued in
Ref. \cite{KDuenwegP}. Another possibility is to dampen the motion of
the particles by friction and compensate this via stochastic thermal
noise (Langevin simulation, so-called Stochastic Dynamics). The
difference in dynamics would then be implemented via different
friction constants. One disadvantage of this method, though, is that
it does not conserve the momentum, and hence does not describe the
hydrodynamics correctly\cite{Duenweg93,Duenweg_Alb}. The latter might
however be important for the phenomenon. Fortunately, there exists a
variant of Stochastic Dynamics, called Dissipative Particle Dynamics
(DPD)\cite{Espanol,thoso}, which does not suffer from this
deficiency. In this method only {\em relative} velocities are dampened
so that Galilean invariance is restored. Furthermore, the stochastic
force acts on {\em pairs} of particles so that Newton's third law
(momentum conservation) holds. For more details, see
Sec. \ref{sec:DPD}. We have hence introduced the difference in dynamic
properties via suitable DPD friction coefficients $\zeta_{ss}$,
$\zeta_{sf}$, and $\zeta_{ff}$. Here, the indices ``s'' and ``f''
stand for ``slow'' and ``fast'', respectively. A friction coefficient
$\zeta_{\alpha \beta}$ indicates the rate of damping of the relative
velocity between two particles of species $\alpha$ and $\beta$,
respectively. To our knowledge such a ``binary generalization'' of
DPD simulations, though fully straightforward, has not yet been
considered. The advantage of this approach is that such a model is
more flexible than just implementing different masses, since it allows
for three parameters rather than just two single-species
parameters. In particular, it is possible to model the ``plasticizing''
effect in a very simple way by choosing a small value for
$\zeta_{sf}$ (as soon as a slow particle is in a fast environment,
its motion is no longer dampened).

In our data analysis, we focus on the composition profiles indicative
of the interdiffusion. Moreover, instead of just looking at the time
scaling behavior of the position of the ``interface'' (that is, of the
locus of a given composition), we rather compare the profiles (or,
more precisely, their Fourier components) directly to the solution of
the (simple) diffusion equation. Fickian and non-Fickian behavior is
then detected by determining the scaling of various quantities with
time.

The remainder of this article is organized as follows: Section
\ref{sec:DPD} gives a brief outline of the standard DPD method.
Section \ref{sec:model} then describes our generalization of the
established version to the binary case with three friction parameters,
and specifies the simulation model in detail. Two versions of the
model are studied: While Model I exhibits pure Fickian diffusion,
Model II reveals anomalous Case II behavior. These results are
presented in Sec. \ref{sec:results}. Finally, Sec. \ref{sec:conclus}
concludes with a brief summary.

\section{Dissipative Particle Dynamics}
\label{sec:DPD}
The equations of motion in the DPD algorithm are
\begin{equation}
\frac{d}{dt} \vec r_i  =  \frac{1}{m_i} \vec p_i \quad \quad \quad
\frac{d}{dt} \vec p_i  =  \vec F_i + \vec F^{(fr)}_i + \vec F^{(st)}_i
\end{equation}
where $\vec r_i$ is a particle position, $\vec p_i$ is a particle momentum, 
and $m_i$ is the mass of particle $i$.  Defining $\vec r_{ij} = \vec r_i - 
\vec r_j = r_{ij} \hat r_{ij}$, (here $\hat r_{ij}$ denotes the respective 
unit vector), we can write the conservative force as
\begin{equation}
\vec F_i = \sum_j -(d{U}_{ij}/dr_{ij}) \hat r_{ij}
\end{equation}
that satisfies Newton's third law (here $U_{ij}$ is the interaction potential 
between particles $i$ and $j$). The friction force $\vec F^{(fr)}_i$ is 
\begin{equation}
\vec F^{(fr)}_i = - \sum_j \zeta(r_{ij})
\left[ \left( \vec v_i - \vec v_j \right) \cdot \hat r_{ij} \right]
\hat r_{ij},
\end{equation}
where $\zeta$ is the friction coefficient. Similarly, we get the
stochastic forces along the inter-particle axes:
\begin{equation}
\vec F^{(st)}_i =  \sum_j \sigma(r_{ij}) \, \eta_{ij} (t) \, \hat r_{ij},
\end{equation}
where $\sigma$ is the noise strength related to the friction
coefficient $\zeta$ through the Fluctuation-Dissipation Theorem (FDT)
$\sigma^2(r) = k_B T \zeta(r)$ and $\eta_{ij}(t)$ is a Gaussian white
noise variable with the properties $\eta_{ij} = \eta_{ji}$,
$\left< \eta_{ij} \right> = 0$, and $\left< \eta_{ij} (t)
\eta_{kl} (t^\prime) \right> = 2 ( \delta_{ik} \delta_{jl} +
\delta_{il} \delta_{jk} ) \delta(t - t^\prime)$. It is then easy
to see that the relations 
\begin{equation}
\sum_i \vec F^{(fr)}_i =
\sum_i \vec F^{(st)}_i = 0
\end{equation}
hold, i.~e. momentum is conserved.

\section{The Model}
\label{sec:model}

The DPD technique allows to construct a simple model of a binary
system of slow and fast particles. As DPD makes use of friction
parameters to control the momentum relaxation and the mobility of the
particles, it is possible to choose the two types of beads to be
completely identical as far as their static properties (masses, sizes,
and interactions) are concerned. The simplifying aspect of such a
model is that its equilibrium properties are trivially known - the
equilibrium state is just a random mixture.

The equations of motion are similar to those in the standard DPD
algorithm. Only the friction and the stochastic terms are slightly
modified as follows:
\begin{equation}
\vec F_{ij}^{fr} = - \zeta_{\alpha \beta} (r_{ij})
\left[ \left( \vec v_i - \vec v_j \right) \cdot \hat r_{ij} \right]
\hat r_{ij}
\end{equation}
\begin{equation}
\vec F_{ij}^{st} = \sigma_{\alpha \beta}(r_{ij}) \, 
\eta_{ij} (t) \, \hat r_{ij}
\end{equation}
As functions $\zeta_{\alpha \beta} (r)$ and $\sigma_{\alpha \beta}(r)$
we simply take constant values
up to a certain cutoff for which we choose the same value $r_c$
as for the interaction potential (see below). The mutual
friction coefficients $\zeta_{\alpha \beta}$ can take 3 different
values: When two slow particles interact $\zeta_{\alpha
\beta} = \zeta_{ss}$, when two fast particle interact 
$\zeta_{\alpha \beta} = \zeta_{ff}$, while $\zeta_{\alpha \beta} =
\zeta_{sf}$ when a slow and a fast particle interact. $\sigma_{\alpha
\beta}$ is related to $\zeta_{\alpha \beta}$ according to the
fluctuation-dissipation theorem: $\sigma_{\alpha \beta} = (2 k_B T
\zeta_{\alpha \beta})^{1/2}$.

The interaction potential $U_{ij}$ is chosen as a repulsive
Lennard-Jones potential between particles $i$ and $j$:
\begin{equation}
U_{ij}(r_{ij}) = 
\begin{cases}
4 \epsilon [ (\frac{\sigma^{LJ}}{r_{ij}})^{12} - 
 (\frac{\sigma^{LJ}}{r_{ij}})^{6} + \frac{1}{4}], &  r_{ij} < r_c \\ 
 0, &  r_{ij} \ge r_c 
\end{cases}
\end{equation}
where $r_c = 2^{1/6}\sigma^{LJ}$ is the cutoff radius for all the forces
acting on bead $i$. The unit system is defined by setting the
particle masses, the energy parameter $\epsilon$, and the
length $\sigma^{LJ}$ to unity.

The initial equilibrium configurations of such a system are prepared
in a rectangular box ($L_x \times L_y \times L_z$, where $L_x = L_y = 36$
and $L_z = 18$) with periodic boundary conditions (pbc) in $x$,
$y$ and $z$ directions. At time $t=0$ a slab spanning half the box ($-
L_x / 4 < x < L_x / 4$) is defined and the property ``slow'' is
assigned to each particle within that slab. The other particles are
defined as ``fast''. Due to the pbc the fast particles (solvent) penetrate
the region of slow particles (matrix) from both sides and thus
create two fronts.

From this moment on the mixing process is monitored by measuring the
density profiles of the slow and of the fast particles.  The values we
choose for the $\zeta_{\alpha \beta}$
are: 1) $\zeta_{ss} = 10$, $\zeta_{sf} =
1.0$ and $\zeta_{ff} = 0.1$ (system I), time step $\delta t = 2 \times
10^{-3}$ for the integration of the equations of motion, and 2)
$\zeta_{ss} = 1000$, $\zeta_{sf} = 1$ and $\zeta_{ff} = 1$ (system II)
whereby the large friction coefficient $\zeta_{ss} = 1000$ requires a
very small time step: $\delta t = 5 \times 10^{-5}$. The number of
particles in the box is $2 \times 10^{4}$ at total density $\rho =
0.85$ which is kept constant throughout the simulation. The starting
configurations are fully equilibrated by means of DPD simulations
using the friction parameter $\zeta_{ff}$. All the simulation
experiments are held at temperature $k_BT = 1.0$.  We measure the
profiles every $10^3$ MD steps for system I and every $10^4$ for
system II. 70 independent runs are made for system I up to a time
$t=2500$ ($= 1.25\times 10^6$ MD time steps) where the final stage of
the mixing process is established. The results obtained for system II
are averaged over 6 independent runs up to time $t = 1725$ ($ = 3.45
\times 10^7$ MD time steps) where the system is still far away from an
equilibrium mixture. Note that the time step $\delta t$ should be
carefully chosen since a too large integration step may strongly
influence the numerics and lead to unphysical results. To this end we
have performed additional control runs using a five times smaller
integration step and verified that our control results do not deviate
from those which were derived with the reported $\delta t$. We show a
comparison of these at the end of the following section.

\section{Simulation Results}
\label{sec:results}\

\begin{figure}
\vspace*{6.0cm} 
\includegraphics{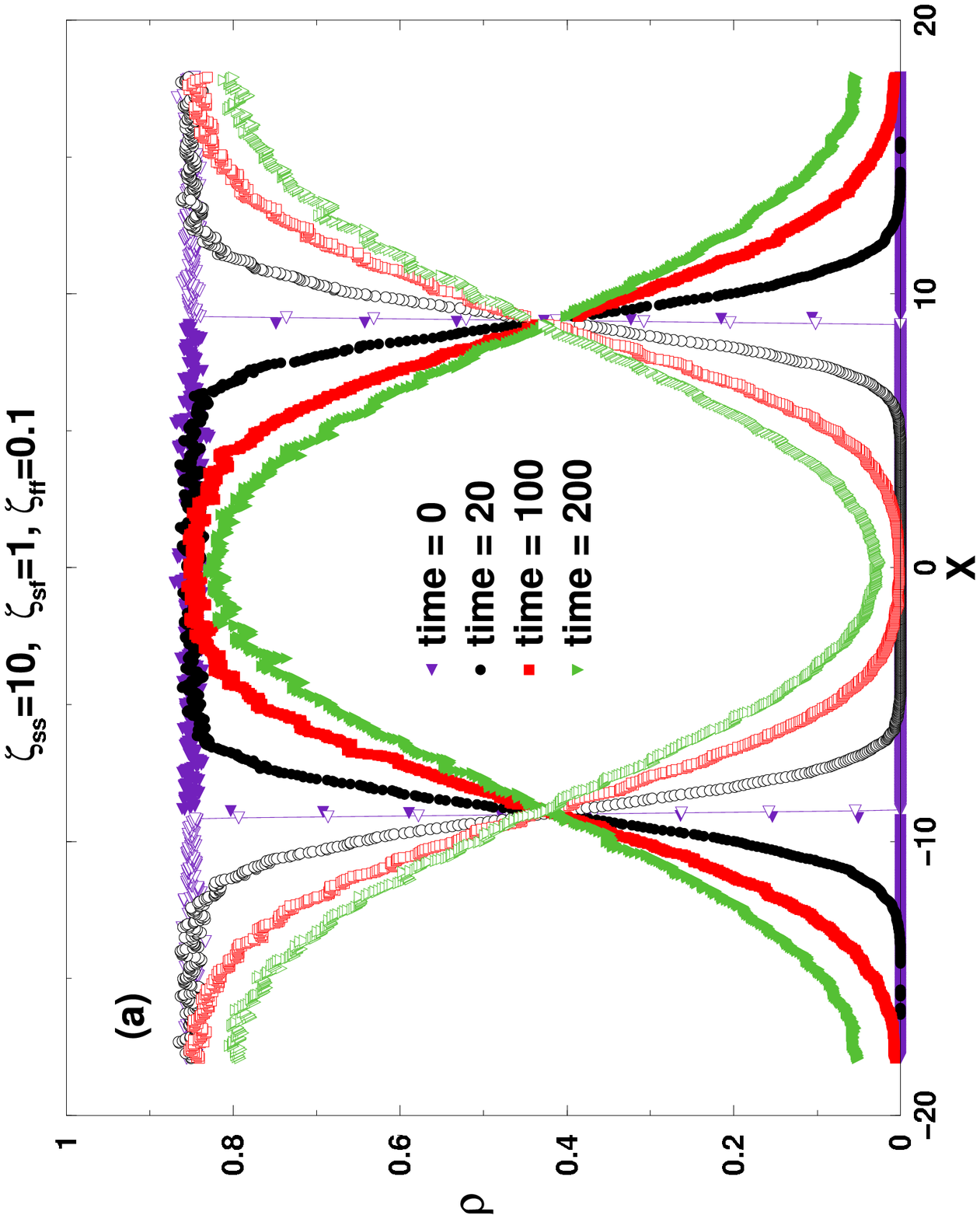}
\includegraphics{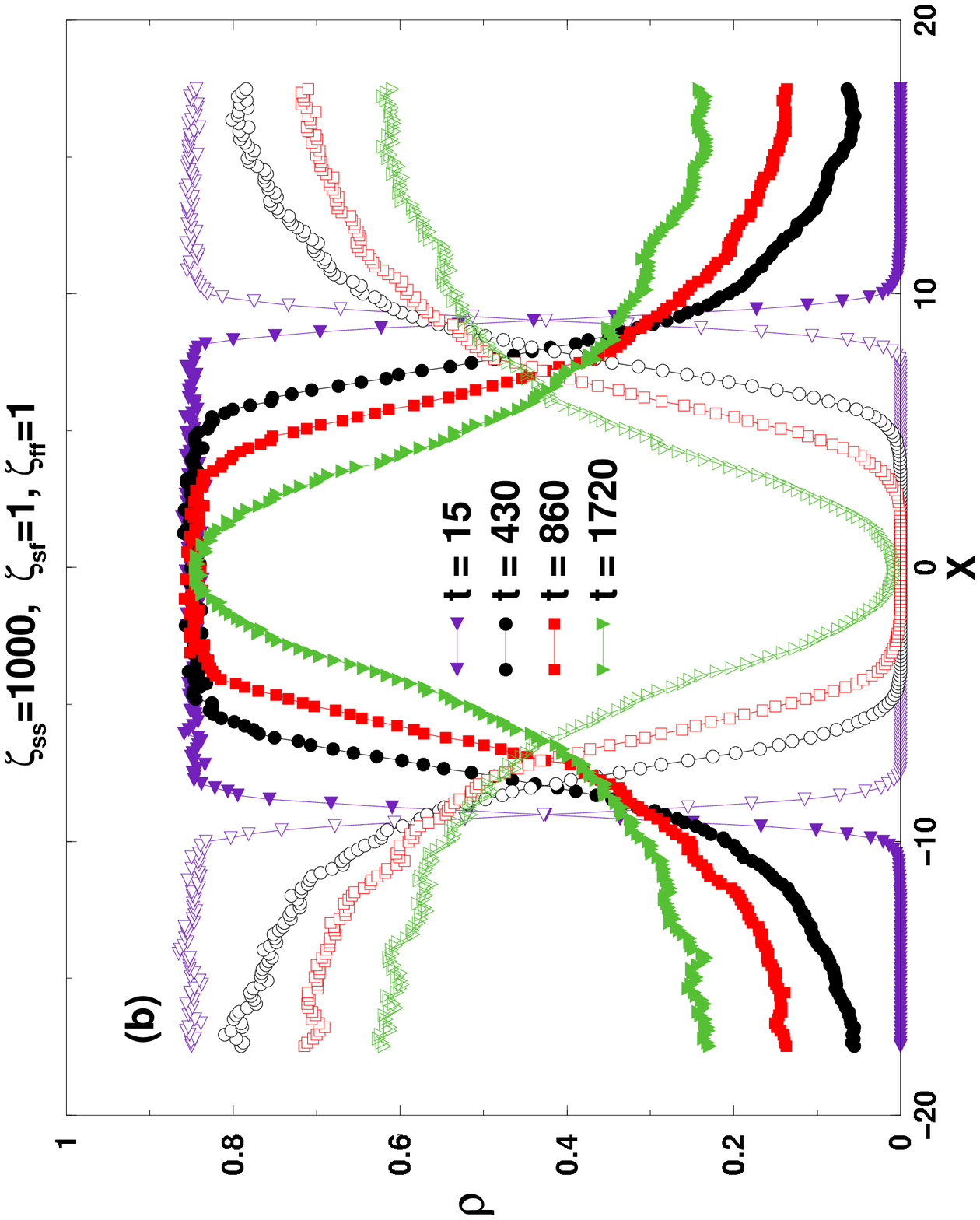}
\vspace*{14.0cm} 
\caption{\label{fig1}}
\end{figure}

Density profiles obtained by using the adopted simulation method are
shown in Fig.\ref{fig1}. Principally two processes take place during
matrix dissolution: 1) the establishment of a smeared-out front as a
result of Fickian diffusion, and 2) movement of the front
position. Case II diffusion is observed when the second process is
much faster than the first one\cite{Taylor00}. This regime is achieved
when the difference in the tracer diffusion coefficients of slow
particles in the matrix $D_s^m$ and in the solvent $D_s^s$ is large
enough, i.~e. $D_s^s \gg D_s^m$. This is the assumption of
Ref. \cite{Taylor00} while the effect of ``plasticizing'' in Rossi's
model\cite{deGennes95} is implemented via $D_s^s = \infty$.

The behavior of system I (Fig. \ref{fig1}a) does not follow this
scheme and a deviation from Fickian diffusion is not observed. The
difference in friction coefficients is small and the profile of slow
particles in the fast environment looks essentially the same as that
of the fast particles in the slow matrix. In contrast, a lack of such
symmetry, due to the presence of {\em both} processes, the moulding of
the front {\em and} its moving in time, can be noticed in system II
(Fig. \ref{fig1}b) where a regime of non-Fickian transport is
attained.

\begin{figure}
\vspace*{7cm}
\includegraphics{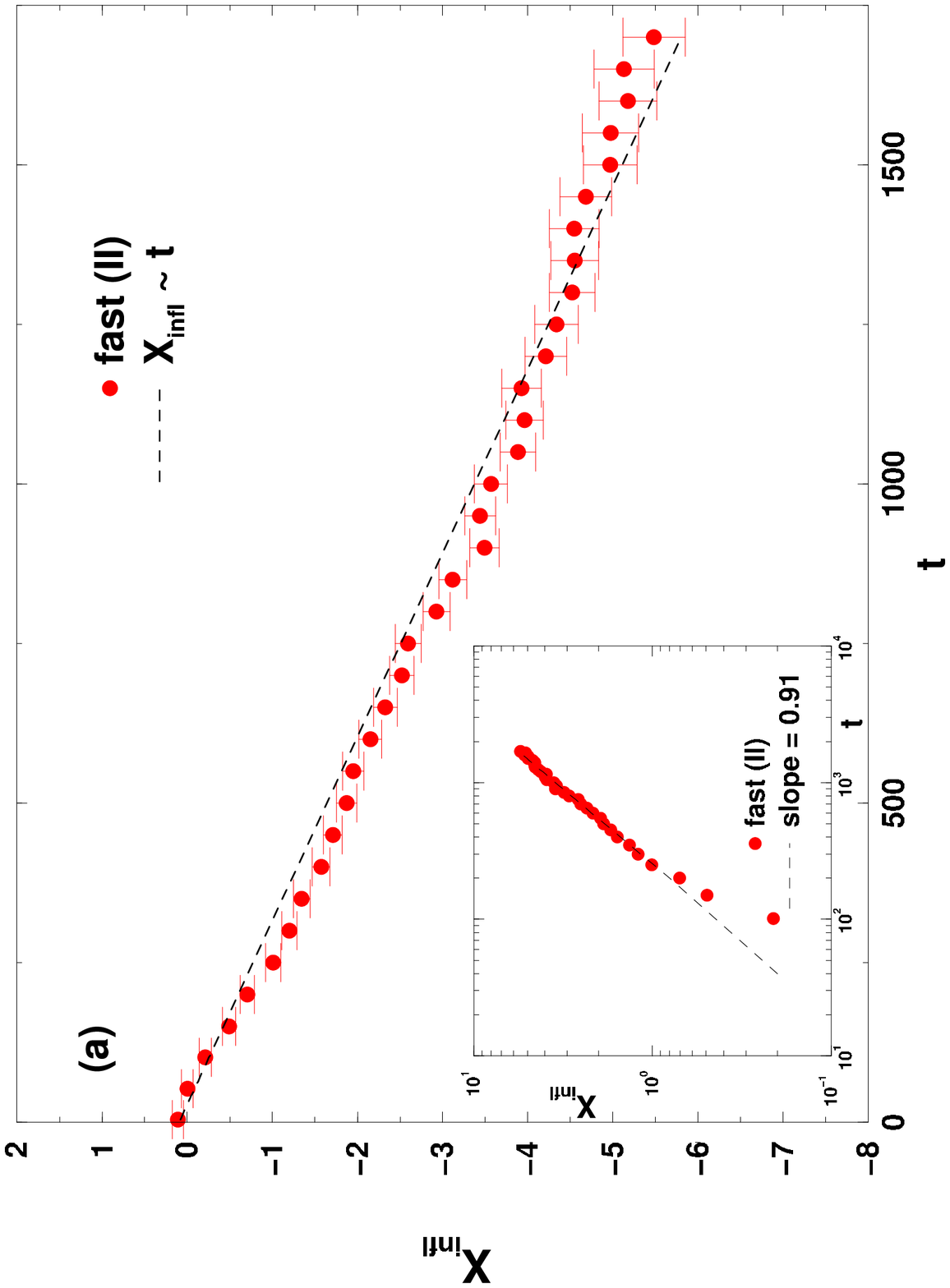}
\includegraphics{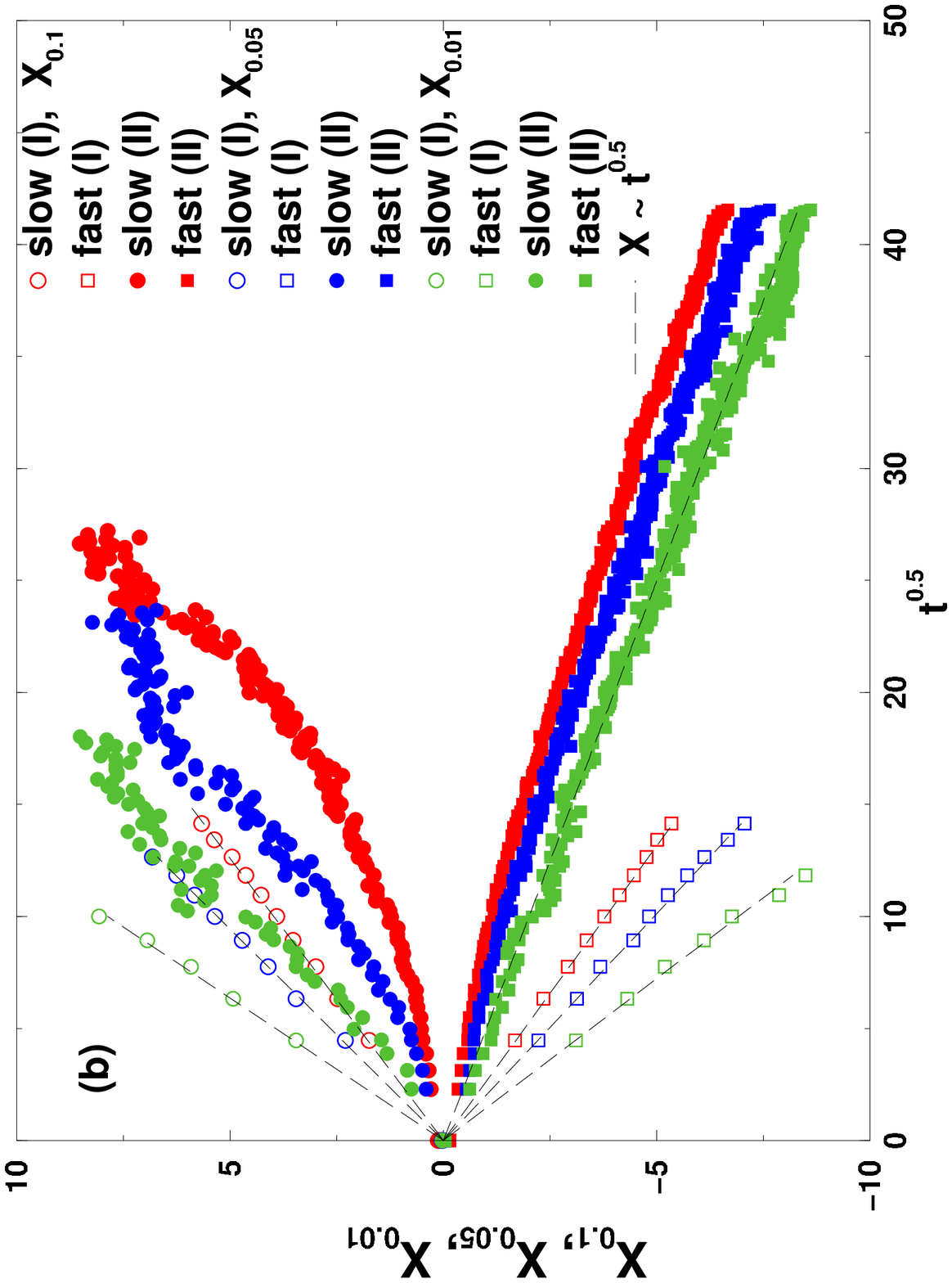}
\vspace*{14cm}
\caption{\label{fig2}}
\end{figure}

The dynamics of penetration of fast particles in the matrix of slow
particles and vice versa can be analyzed by looking at the front
propagation (the position of each front is determined by the
inflection point of the respective density profile), Fig. \ref{fig2}a,
and at the trajectories of points with constant density,
Fig. \ref{fig2}b. The motion of the front, $X_{infl}$, is almost
negligible in system I (not shown in Fig. \ref{fig2}a), and only a slight
deviation from its constant value at early times can be detected. Its
velocity is much smaller than the speed of the process of front shape
formation which governs the behavior of the system.

A nearly linear dependence of $X_{infl}(t)$ is observed in system II
up to the latest time of the simulation, as expected for Case II
diffusion. However, also a continuous decay of the front velocity is
detected. One could expect that after time $t \approx 1300$ a
crossover from Case II to Case I behavior develops and the system
goes back to Fickian diffusion.

The three positions of constant density, $X_{0.1}(t)$, $X_{0.05}(t)$
and $X_{0.01}(t)$ in system I (Fig. \ref{fig2}b) follow the typical
behavior of Fickian diffusion\cite{Grest04,Grest04_1} $X(t) \propto
t^{0.5}$. On the other hand, a crossover from conventional to
anomalous diffusion shows up in the data of system II as the points of
constant concentration (for which we measure the propagation of
density profiles) change to higher values. The motion of the point of
lowest concentration in system II, $X_{0.01}$, goes as $t^{0.5}$, and
it defines the depth of the Fickian tail ahead of the moving front. A
slowing-down of the front velocity with time is due to the influence
of this Fickian precursor as argued by Taylor\cite{Taylor00}.

Results obtained by a Fourier analysis of the density profiles are shown
in Fig. \ref{fig3}. Starting from Fick's equation
\begin{equation}
\frac{\partial}{\partial t} C(x,t) = D \frac{\partial}{\partial x^2} C(x,t) ,
\end{equation}
where $C(x,t)$ is the density profile in x-direction and $D$ is the
collective diffusion coefficient, the concentration $C(x,t)$ is
represented as a Fourier series ($L \equiv L_x$)
\begin{eqnarray}
C(x,t) & = & \sum_{p=0}^{\infty} C_p (t) \cos 
             \left( \frac{2 \pi p x}{L} \right) \\
C_p(t) & = & \frac{2}{L} \int_{-L/2}^{L/2} dx \, C(x,t) \cos
             \left( \frac{2 \pi p x}{L} \right) \quad p > 0 \\
C_0(t) & = & \frac{1}{L} \int_{-L/2}^{L/2} dx \, C(x,t) .
\end{eqnarray}
This representation has the advantage that both the periodic boundary
conditions and the finite system size are automatically taken into
account. Then the equation is solved with respect to the Fourier
coefficients $C_p$, resulting in $C_p(t) = C_p(0) \exp (- D (2 \pi p /
L)^2 t)$.  It can be shown that $C_p$ for even $p$ is always $0$ as a
consequence of our rectangular-shaped initial condition; hence we
consider here only the odd coefficients. The normalized Fourier
coefficients $\frac{C_p(t)}{C_p(0)}$ plotted vs scaled time $(2 \pi p
/ L)^2 t$ should collapse on a master curve (exponential decay) if the
process was a normal diffusion. The results in Fig. \ref{fig3}a follow
exactly this behavior and the collective diffusion coefficient
estimated from the decay is $D \approx 0.054$. Note that using the
argument $t / L^2$ implies the Fickian scaling $X \sim t^{1/2}$.

In contrast, the behavior of system II (Fig. \ref{fig3}b) reveals a
strong deviation from the above scenario due to the moving front: The
data clearly do not collapse on a single curve (indicating a violation
of Fickian scaling), and overshoot into negative values. In the inset,
a linear scaling $X \sim t$ is assumed.  This scaling is better
although not perfect. Furthermore, it is seen that the reduction of
the $L$ exponent ``overcorrects'', i.~e. the best collapse would be
observed for an exponent somewhere between 1 and 2. Hence, the data
seem to indicate that the system is in a crossover region somewhere
between the limiting Cases I and II.

In order to understand what one should expect for the behavior of the
Fourier coefficients in the limiting Case II, we calculated them for a
simple model where a sharp moving front is the only feature of the
system. We assume that two sharp fronts approach each other with
constant velocities $\pm v$ whereby the profile retains a rectangular
shape. One then obtains for the odd coefficients $C_p(t) = C_p(0)
\cos(2 \pi p v t / L)$. Of course, this result exhibits linear scaling
$X \sim t$. Although the corresponding data collapse is not perfect,
this result qualitatively explains the overshoot into the negative
regime, which, according to the analytical result, should occur for
all modes $p \ge 3$. One should keep in mind that the front in system
II is not a vertical line but has a more complex shape (where
different parts of the profile follow different laws of motion) and
the velocity of the front is not constant throughout the simulation
but decreases with time. For these reasons, the coefficients show a
more complex behavior.

We also tried another type of simulation where the velocity of the
front is kept constant during the simulation by artificially removing
the tails of the profiles: immediately after a slow particle is
dissolved and is surrounded only by fast particles (the detection of
such an event being part of the DPD procedure anyway), we remove it
from the system and put a fast particle in its place. This model
attempts to mimic the process of matrix dissolution where the slow
particles precipitate after they split off from the surface. The set
of friction coefficients is the same as in system II. In this case
the density profile of slow particles looks like a step-function with
height $\rho$ and erfc-like boundaries. The front moves linearly with
time. The Fourier coefficients plotted vs $(2
\pi p / L) t$ (data not shown) are then cosine waves with same
periods (the whole profile shrinks with constant velocity) but
different amplitudes (due to the erfc-like shape).

The regime of anomalous diffusion in system II is reached due to the
difference in momentum relaxation of slow particles in the matrix and
in the solvent. In order to check the influence of the dynamics of
fast particles on the process of dissolution we also use another set
of friction coefficients: $\zeta_{ss} = 1000$, $\zeta_{sf} = 1$ and
$\zeta_{ff} = 0.001$. The results (data not shown) are similar to
those obtained in system II. This is expected, since even in the limit
of vanishing friction the interactions in the dense repulsive
Lennard-Jones fluid provide a particle friction of roughly $\zeta =
20$ \cite{KDuenwegP}. Whenever $\zeta_{ff}$ is small compared to
this value, the dynamics of the fast particles is dominated
by the conservative forces.

Eventually, in Fig. \ref{fig4} we compare our data for the Fourier
coefficients $C_p(t)/C_p(0)$ in the case of non-Fickian diffusion to
data obtained by using a five times smaller step of integration
$\delta t$. Clearly, the results are consistent with one
another within the limits of statistical accuracy (the curves with the
smaller $\delta t$ value were obtained from a single run). This
demonstrates that the chosen parameters of the MD simulation do not
introduce any distortion of the physical picture.

\section{Conclusions}\label{sec:conclus}

We have constructed a very simple particle-based computer simulation
model which is able to exhibit substantial deviations from Fickian
diffusion, and shows a clear crossover towards Case II behavior
(i.~e. a constant-velocity front with essentially flat concentration
profiles ahead and behind). The model disregards all molecular details
but incorporates the two essential features which we consider as
necessary for reaching the Case II limit: A substantial discrepancy in
the mobility of the two (pure) species, and a ``plasticizing'' effect,
i.~e. a substantial increase in the mobility of the slow species as
soon as its molecules are surrounded by the fast species. Our
simulation results further corroborate the view that no additional
molecular ingredients are necessary. We have no definitive explanation
why the simulations by Tsige and Grest\cite{Grest04,Grest04_1} were
unable to reach the Case II limit, but speculate that the amount of
plasticization might have been insufficient.

What we find remarkable about our results is not so much the observed
physics as such -- the basic mechanism is essentially rather simple,
and the Case II limit should be expected as soon as the dynamic
asymmetry, combined with plasticization, is strong enough -- but
rather the computational ease with which our model is able to obtain
the data, which are rather close to the desired limit. The model aims
directly at controlling dynamic properties, without modifying the
equilibrium statics. Furthermore, since we have three parameters
($\zeta_{ss}$, $\zeta_{ff}$ and $\zeta_{sf}$) at our disposal,
plasticization is particularly straightforward to implement by
choosing a small value for $\zeta_{sf}$. An additional bonus is that
hydrodynamics (momentum conservation) is taken faithfully into
account. We believe that our approach via DPD is particularly
promising for simulations of Case II systems. For example, it should
now be possible to combine our approach with a polymer simulation.
There the DPD part would model the ``glassy'' (slow) behavior of the
matrix, plus plasticization, while the modeling in terms of polymer
chains would account for the connectivity and entanglements. Thus
interesting questions of molecular stretching, reorientation,
disentanglement kinetics, stresses, etc., which are difficult to treat
in terms of macroscopic models, could be attacked.

\section{Acknowledgments}

One of us, J. Y. is indebted to the MPIP, Mainz for hospitality and
assistance during her stay in this institution, and to the M. Curie
training site for financial support.


\clearpage
\section*{Figure captions}

\begin{figure}
\vspace*{5.5cm}
\includegraphics{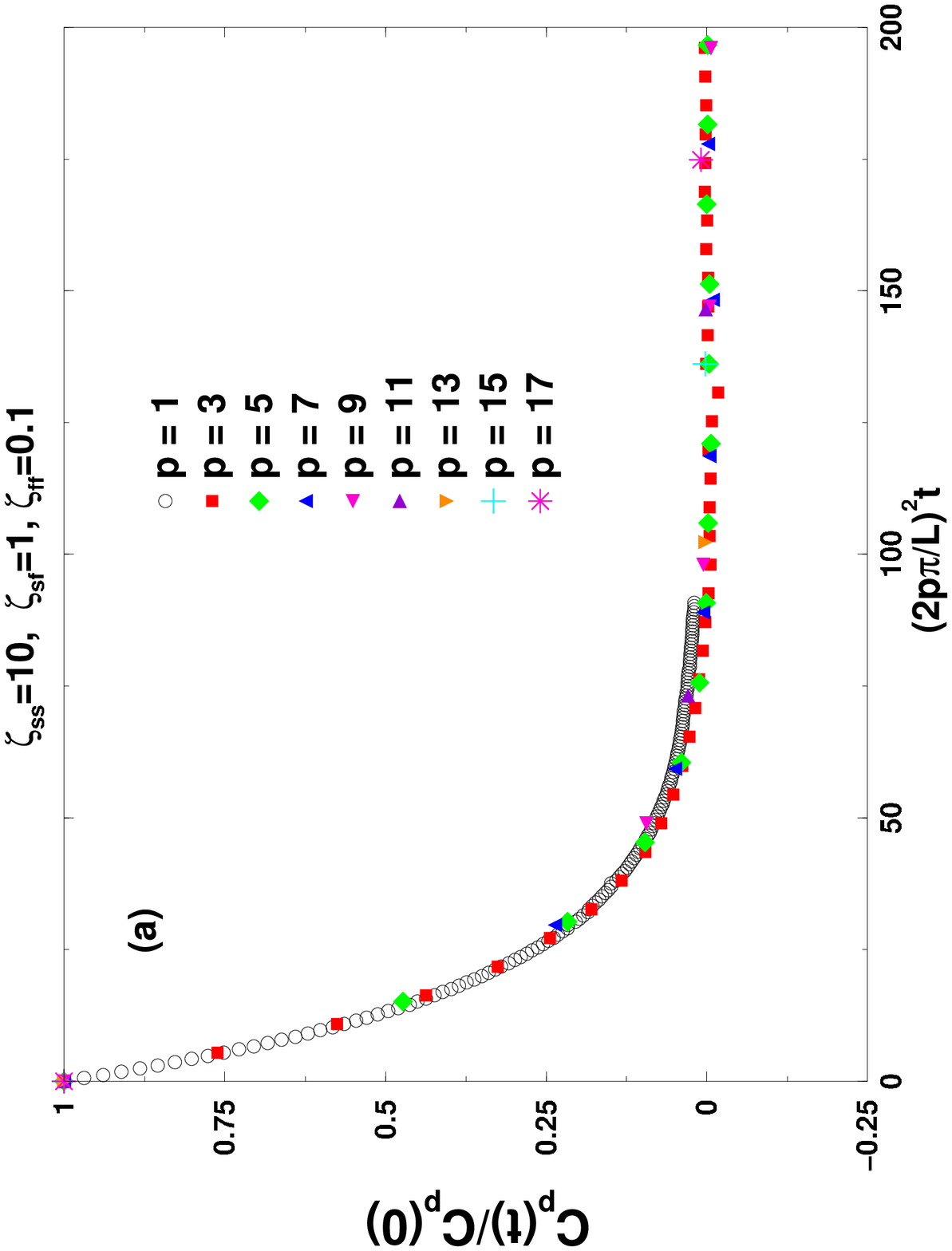}
\includegraphics{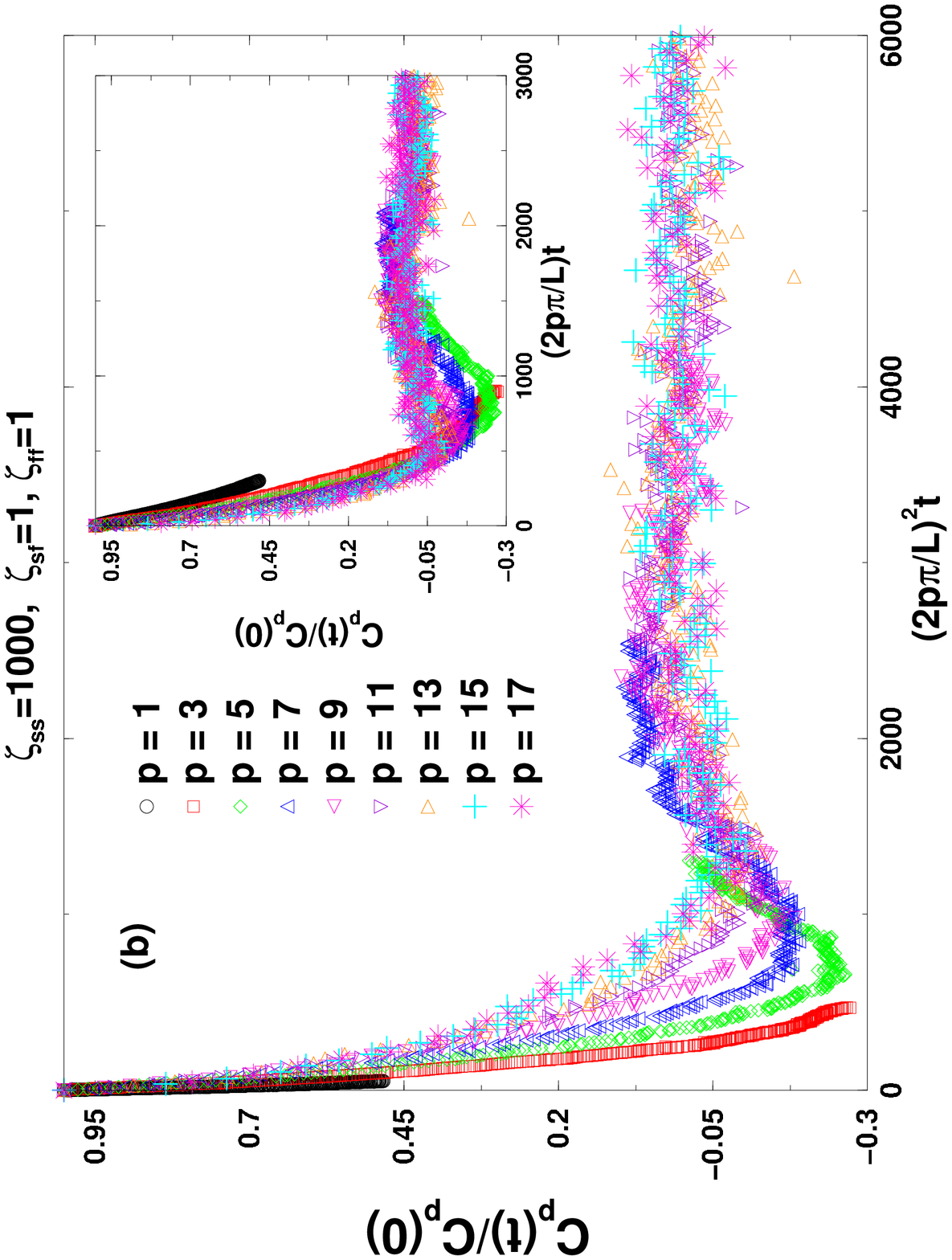}
\vspace*{14cm}
\caption{\label{fig3}}
\end{figure}

\begin{figure}
\vspace*{6cm}
\includegraphics{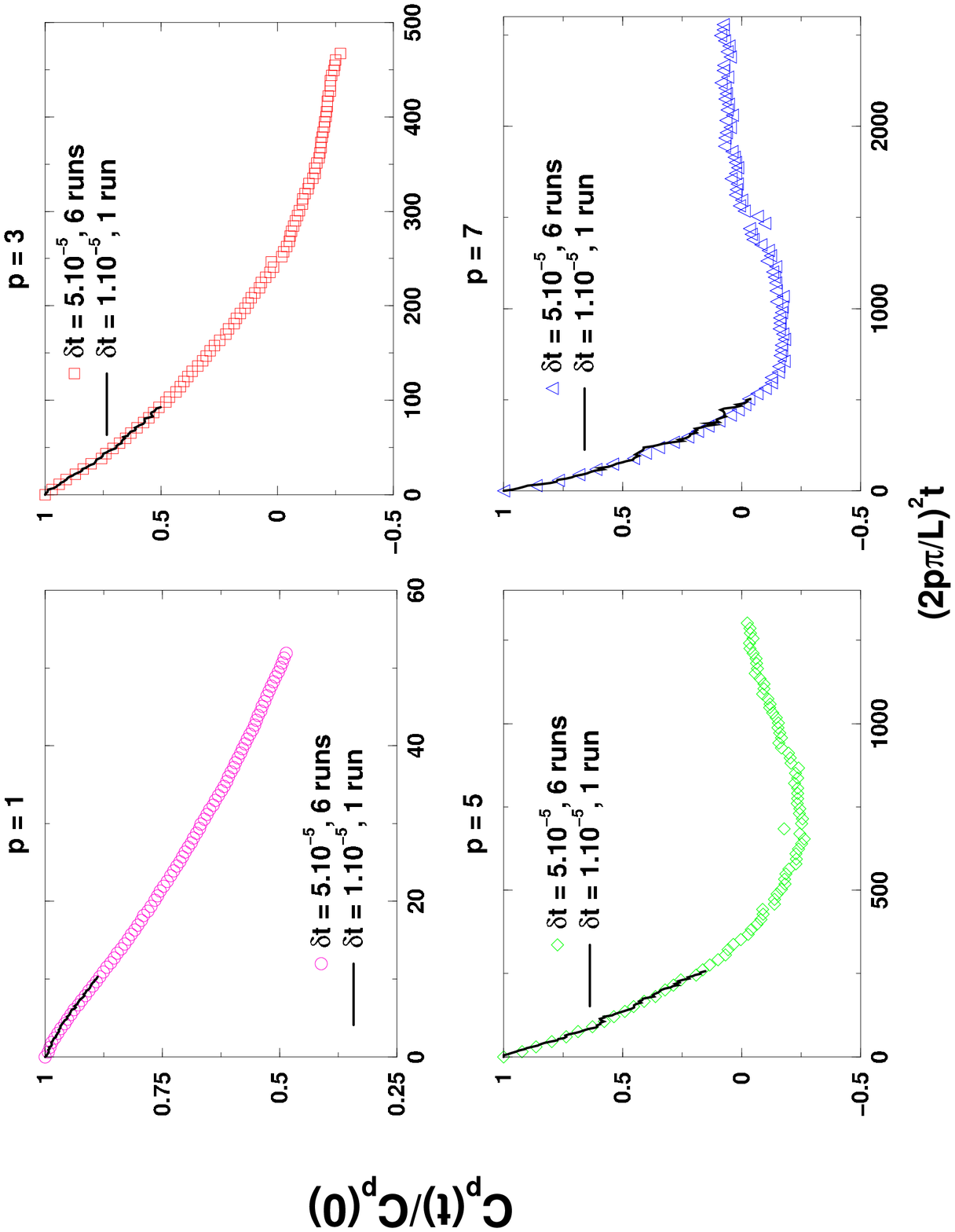}
\vspace*{7cm}
\caption{\label{fig4}}
\end{figure}

\begin{enumerate}
\item
Density profiles in x-direction of the slow particles (full symbols)
and of the fast particles (open symbols) for 4 different
times. Part (a): system I, part (b): system II.

\item
(a) Motion of the front position $X_{infl}$ of fast particles for
system II. The dashed line denotes
the dependence characteristic for Case II diffusion $X(t) \propto
t$. The data are shifted so that their initial position is in the
middle of the box $X(t=0.0) = 0.0$.  A log-log plot of $X_{infl}$ of
system II is shown in the inset of part (a) where the slope (dashed
line) is $0.91 \pm 0.01$. Three positions of constant density $\rho_1
= 0.1$ ($X_{0.1}$), $\rho_2 = 0.05$ ($X_{0.05}$) and $\rho_3 = 0.01$
($X_{0.01}$) are plotted in (b) against square root of time for the
slow particles (circles) and for the fast particles (squares). 
Open symbols denote the data of system I, full symbols those
of system II. The results are shifted as in part
(a). Dashed lines in (b) mark the dependence of $X(t)$ typical for
Fickian diffusion, $X(t) \propto t^{0.5}$.

\item
The first 9 Fourier coefficients $C_p$ ($p = 1, 3, 5 \ldots 17)$ of
the density profiles vs $(2 \pi p / L)^2 t$. Part (a): system I, part
(b): system II. Inset of part (b): The first 9 Fourier coefficients
$C_p$ ($p = 1, 3, 5 \ldots 17)$ of the density profiles vs $(2 \pi p /
L)t$.

\item
The same as in Fig. \ref{fig3}b whereby symbols denote data obtained with 
$\delta t = 5 \times 10^{-5}$ while full lines are obtained with 
$\delta t = 10^{-5}$.

\end{enumerate}

\clearpage

\clearpage

\clearpage

\clearpage

\end{document}